\documentclass[]{pasj02} 
\usepackage[switch,mathlines]{lineno} 
\usepackage{natbib,bm}

\jyear{2024}
\Received{2024/08/21}
\Accepted{2024/11/18}

\newcommand{\pfrac}[2]{ \biggl(\dfrac{#1}{#2}\biggr) }

\begin{document} 

\title{Modeling protoplanetary disk heating by planet-induced spiral shocks}

\author{
 Tomohiro \textsc{Ono},\altaffilmark{1}\orcid{0000-0001-8524-6939}  
 Tatsuki \textsc{Okamura},\altaffilmark{2} 
 Satoshi \textsc{Okuzumi},\altaffilmark{2}\orcid{0000-0002-1886-0880}\altemailmark\email{okuzumi@eps.sci.titech.ac.jp}
 and 
 Takayuki \textsc{Muto}\altaffilmark{3}
}
\altaffiltext{1}{School of Natural Science, Institute for Advanced Study, 1 Einstein Drive Princeton, New Jersey, 08540, USA}
\altaffiltext{2}{Department of Earth and Planetary Sciences, Institute of Science Tokyo, Ookayama, Meguro-ku, Tokyo, 152-8551, Japan}
\altaffiltext{3}{Division of Liberal Arts, Kogakuin University, 1-24-2 Nishi-Shinjuku, Shinjuku-ku, Tokyo 163-8677, Japan}




\maketitle

\begin{abstract}
We investigate the heating of protoplanetary disks caused by shocks associated with spiral density waves induced by an embedded planet. Using two-dimensional hydrodynamical simulations, we explore the dependence of shock heating rates on various disk and planetary parameters. Our results show that the shock heating rates are primarily influenced by the planet's mass and the disk's viscosity, while being insensitive to the {thermal relaxation} rate and the radial profiles of the disk's surface density and sound speed. We provide universal empirical formulas for the shock heating rates produced by the primary and secondary spiral arms as a function of orbital radius, viscosity parameter $\alpha$, and planet-to-star mass ratio $q$. {The obtained formulas are accurate within a factor of a few for a moderately viscous and adiabatic disk with a planet massive enough that its spiral arms are strongly nonlinear.} Using these universal relations, we show that shock heating can overwhelm viscous heating when the disk viscosity is low ($\alpha \lesssim 10^{-3}$) and the planet is massive ($q \gtrsim 10^{-3}$). Our empirical relations for the shock heating rates are simple and can be easily implemented into radially one-dimensional models of gas and dust evolution in protoplanetary disks.
\end{abstract}


\section{Introduction}\label{sec1:Intro}

Protoplanetary disks are the birthplaces of planets. Their physical properties, such as temperature and ionization structures, influence the characteristics of the planets that form within them. The disk's temperature structure is of particular interest because it plays a crucial role in determining the chemical composition of forming planets \citep[e.g.,][]{Bitsch19}. For instance, the location of the snow line significantly influences the water content of protoplanets \citep[e.g.,][]{Sato16}, although other internal processes can also affect their water budgets \citep{Lichtenberg19, Johansen21}. 

Most models of disk thermal evolution \citep[e.g.,][]{Bitsch2015, Mori2021} assume that disks are primarily heated by stellar irradiation and/or accretion heating. However, this assumption may not hold for disks that already harbor a massive planet. A planet embedded in a gas disk excites spiral density waves \citep[e.g.,][]{GoldreichTremaine1979}, which eventually shock as they propagate away from the planet \citep{GoodmanRafikov2001, Rafikov2002}. These shocks generate entropy, causing secular heating of the disk gas \citep{Rafikov2016}. If the planet is massive and the disk is optically thick, this shock heating can raise the disk temperature \citep{Richert2015, Lyra2016, Rafikov2016, Ziampras2020}, potentially altering the chemical composition of the gas and dust remaining in the disk. Recent discoveries of protoplanetary disk substructures \citep[e.g.,][]{Andrews2020, Bae23} and isotopic dichotomies in meteorites \citep[e.g.,][]{Kruijer20, Kleine20} suggest that giant protoplanets may have formed while their parent disks, including the solar nebula, were still abundant in solids. This implies that planet-induced shock heating could have impacted the composition of solar system bodies and exoplanets that formed from solids in the vicinity of giant planets.

Recently, \citet{Ziampras2020} conducted a systematic study of planet-induced shock heating using two-dimensional (2D) hydrodynamic simulations of disks with an embedded giant planet. They demonstrated that shock heating by a Jupiter-mass planet can significantly alter the disk's temperature profile and even shift the location of the snow line. However, the disk temperature obtained directly from such simulations depends not only on shock heating but also on the disk's radiative cooling (i.e., opacity) and accretion heating. Both the disk's opacity and accretion heating rate are largely uncertain and can change with dust evolution \citep[e.g.,][]{Oka_2011, Kondo23, Delage23, Fukuhara24}. To model planet-induced shock heating as independently as possible from these uncertainties, it is necessary to extract the shock heating rate from these simulations and analyze its dependence on the planet's mass and distance from the planet.

In this paper, we model the planet-induced shock heating rate using 2D hydrodynamic simulations of disks with an embedded planet. From these simulations, we extract the radial profiles of the shock heating rates by directly measuring the jumps in the disk's specific entropy across planet-induced shock waves. We discover universal scaling relations for the radial heating rate profiles from the simulations and provide simple analytic expressions for these universal relations.

The structure of this paper is as follows.
In Section \ref{sec:syuhou}, we describe the simulation setups and analysis methods. Section \ref{sec:kekka} presents our simulation results and analysis. Section \ref{sec:giron} discusses an application and caveats of our study, and Section \ref{sec:Sumally} presents a summary.

\section{Method}\label{sec:syuhou}

\subsection{Basic equations}
\label{subsec:basiceqn} 

We consider a 2D gas disk surrounding a central star of mass $M_\ast$ and a planet of mass $M_\mathrm{p}$ in a circular orbit with an orbital radius of $r_{\mathrm{p}}$. We adopt a frame centered at the position of the central star and co-rotating with the planet. The angular velocity of the planet in the inertial frame is $\Omega_\mathrm{p} = \sqrt{GM_\ast/r_\mathrm{p}^3}$, where $G$ is the gravitational constant. In the co-rotating frame, the planet is located at $ (r, \phi) = (r_\mathrm{p}, 0)$, where $r$ and $\phi$ represent the distance from the central star and the azimuth, respectively.

The equations of continuity and motion in the co-rotating frame are given by
\begin{equation}
    \frac{\mathrm{d}\Sigma}{\mathrm{d} t}=-\Sigma \nabla \cdot {\bm v},
    \label{eq:EoC}
\end{equation}
\begin{equation}
    \Sigma \frac{\mathrm{d} {\bm v}}{\mathrm{d} t} = -\nabla p 
    + {\nabla \cdot \Pi} 
    - \Sigma \nabla {\Phi}
    + \Sigma (\Omega_\mathrm{p}^2{\bm r} 
    - 2\Omega_\mathrm{p}{\bm n}_z \times {\bm v}),
    \label{eq:EoM}    
\end{equation}
where $\Sigma$ is the surface density, ${\bm v} = (v_r, v_\phi)$ is the gas velocity, $\mathrm{d}/\mathrm{d}t = \partial / \partial t + {\bm v} \cdot {\nabla}$ is the 
Lagrangian time derivative, $p$ is the vertically integrated pressure, $\Phi$ is the gravitational potential, ${\bm r}$ is the position vector, and ${\bm n}_z$ is the unit vector perpendicular to the disk plane. 
The components of the viscous stress tensor $\Pi$ are given by 
\begin{equation}
    \Pi_{ij}=\Sigma\nu \left( \frac{\partial v_i}{\partial x_j}
    +\frac{\partial v_j}{\partial x_i}
    -\frac{2}{3}\delta_{ij}{\nabla \cdot \bm v}\right),
\end{equation}
where $\nu$ is the kinematic viscosity. 
We employ the $\alpha$-viscosity model of \citet{Shakura1973} and give $\nu$ as $\nu = \alpha c_{\rm s}H$, where $\alpha$ is the viscosity parameter, $c_{\rm s}$ is the isothermal sound speed, $H = c_{\rm s}/\Omega_{\rm K}$ is the disk scale-height, and $\Omega_\mathrm{K}(r) = \sqrt{GM_\ast /r^3}$ is the Kepler angular velocity. We assume $10^{-4} \leq \alpha \leq 10^{-3}$ (see section~\ref{subsec:parameters}
 for the complete parameter sets). 
The vertically integrated pressure $p$ can be expressed as 
\begin{equation}
    p = c_\mathrm{s}^2\Sigma
    \label{eq:3_p}.
\end{equation}

The equation of energy is given by
\begin{align}
    \frac{\mathrm{d}\Sigma e}{\mathrm{d}t}=& -\gamma \Sigma e {\nabla \cdot{\bm  v}} + \frac{1}{2\nu \Sigma}\left( \Pi_{rr}^2 + 2\Pi_{r\phi}^2+\Pi_{\phi \phi}^2 \right) \nonumber \\
    &+\frac{2\nu \Sigma}{9} ({\nabla \cdot {\bm v}})^2 + Q_\mathrm{relax},
    \label{eq:EoE}
\end{align}
where $e=p/[\Sigma(\gamma-1)]$ is the {specific internal energy (i.e., internal energy per unit mass)}, $\gamma$ is the adiabatic index (fixed to be 1.4 in this study), and $Q_\mathrm{relax}$ is the {thermal relaxation} term. 
We model $Q_\mathrm{relax}$ using the $\beta$-{relaxation (also known as the $\beta$-cooling)} prescription \citep[e.g., ][]{Gammie01}, which enforces $e$ to relax to a reference value on timescale $t_{\rm relax}$ (for details, see section \ref{subsec:num_set}).
{On the right-hand side of Equation~\eqref{eq:EoE}, the first term represents compressional heating and cooling, while the second and third terms account for viscous heating. Due to the disk's differential rotation, the term $\Pi_{r\phi}^2/(\nu\Sigma)$ can, in principle, generate heat even in the absence of a planet. Since our focus is on heating from planet-induced spiral shocks, we subtract the local Keplerian velocity from ${\bm v}$ in the viscous heating terms to exclude heating due to Keplerian shear.}

The net gravitational acceleration ${\bm g}_\mathrm{grav} \equiv - \nabla{\Phi}$ is given by
\begin{align}
    {\bm g}_\mathrm{grav} &= 
    {\bm g}_\star + {\bm g}_\mathrm{p} + {\bm g}_\mathrm{in} \nonumber \\
    &=-\frac{GM_*}{r^3}{\bm r}-\frac{GM_\mathrm{p}}{(r_{\rm e}^2+\epsilon^2)^{3/2}}{\bm r}_{\rm e}-\frac{GM_\mathrm{p}}{r_\mathrm{p}^3}{\bm r}_\mathrm{p}, \\
    {\bm r}_{\rm e}&={\bm r}-{\bm r}_\mathrm{p}, \quad r_{\rm e} = |{\bm r}_{\rm e}| \nonumber
    \label{eq:g}
\end{align}
where ${\bm g}_\star$, ${\bm g}_\mathrm{p}$, and ${\bm g}_\mathrm{in}$ represent the gravitational acceleration by the star, planet, and the indirect term, respectively, and ${\bm r}_{\rm p}$ is the position vector of the planet. 
To avoid infinitely large acceleration in the vicinity of the planet, we adopt a smoothing length of $\epsilon=0.6r_\mathrm{H}$, where $r_\mathrm{H}=(M_\mathrm{p}/3M_\ast)^{1/3}r_\mathrm{p}$ is the planet's Hill radius.
{
Our choice of the smoothing length corresponds to $\approx 0.8H$ for $q=10^{-3}$. \citet{Muller12} show that choosing $\epsilon = (0.6\textrm{--}0.7)H$ best reproduces the vertically averaged gravitational force from the planet in a three-dimensional (3D) disk in the case of low-mass planets.
More recently, the vertical averaging of the 3D planet potential is discussed by \citet{BrownOgilvie24}, who present the form of 2D gravitational potential of a planet that matches 3D hydrodynamic calculations in the context of Type I migration.
}

\subsection{Numerics}\label{subsec:num_set}

We numerically solve equations (\ref{eq:EoC}), (\ref{eq:EoM}), and (\ref{eq:EoE}) using the {\tt Athena++} code \citep{Stone_2020}.
We employ the Harten--Lax--van Leer--Contact approximate Riemann solver, the second-order piecewise linear method for spatial reconstruction, and the second-order Runge--Kutta method for time integration. 
We also adopt the orbital advection algorithm of \citet{Masset2000}. 

The computational domain covers $[r_\mathrm{in}, r_\mathrm{out}] = [0.4r_\mathrm{p}, 9.88r_\mathrm{p}]$ in the radial direction and the full $2\pi$ in the azimuthal direction. Our fiducial runs use 896 and 1584 cells in the radial and azimuthal directions, respectively. We adopt equal spacing in the azimuthal direction and logarithmic spacing in the radial direction, ensuring that each cell has an aspect ratio of approximately unity.
{We also perform a supplementary, low-resolution simulation using $448\times792$ cells to assess whether the finite resolution of our simulations impacts the entropy jumps at shocks (see section~\ref{subsec:overview}).}

The initial surface density and sound speed, $\Sigma_{\rm i}$ and $c_{\rm i}$, are given by
\begin{equation}
    \Sigma_\mathrm{i}(r) = \Sigma_0 \left(\frac{r}{r_\mathrm{p}}\right)^{\xi}, \label{eq:3_s}
\end{equation}
and
\begin{equation}
    c_\mathrm{i}(r) = c_0\left(\frac{r}{r_\mathrm{p}}\right)^{-\xi /2-3/4}, \label{eq:3_c}
\end{equation}
where $\Sigma_0$ and $c_0$ are the initial surface density and sound speed at $r = r_{\mathrm{p}}$, respectively. For a given initial surface density slope $\xi$, the slope of the initial sound speed profile $-\xi /2-3/4$ is determined to ensure that the viscous disk is initially in a steady state with a radially uniform mass accretion rate of $\dot{M}=3\pi\nu\Sigma$ \citep{Lynden-BellPringle1974}. We adopt $c_\mathrm{0}=0.05v_\mathrm{K}(r_\mathrm{p})$, {where $v_{\rm K}(r)$ is the local Keplerian velocity}. 
Our grid resolves the local disk scale height with at least 12 cells for $r \geq 0.55r_\mathrm{p}$. 

The initial gas velocity ${\bm v}_{\rm i} = (v_{r\mathrm{i}}, v_{\phi \mathrm{i}})$ is given by 
\begin{equation}
    v_{r\mathrm{i}} = -\frac{3\alpha c_\mathrm{i}^2}{2r\Omega_\mathrm{K}},
\end{equation}
\begin{equation}
v_{\phi \mathrm{i}}(r) = r\Omega_\mathrm{i}(r)=\sqrt{v_\mathrm{K}^2(r)-c_\mathrm{i}^2(r)},
\end{equation}
where $\Omega_\mathrm{i}$ is the initial angular velocity of the disk.
The initial radial velocity is consistent with steady accretion with $v_{ r} = -{3\nu}/(2r)$ \citep{Lynden-BellPringle1974}.
The initial azimuthal velocity $v_{\phi \mathrm{i}}$ ensures the balance among the central star's gravity, the centrifugal force, and the pressure gradient force.

The {relaxation} term $Q_\mathrm{relax}$ is given by 
\begin{equation}
    Q_\mathrm{relax} = -\frac{\Sigma(e-e_\mathrm{i})\Omega_\mathrm{i}}{\beta} \label{eq:3-q},
\end{equation}
where $\beta$ is a parameter related to the {thermal relaxation} timescale $t_{\rm relax}$ by $t_{\rm relax} = \beta \Omega_{\mathrm{i}}^{-1}$ and $e_\mathrm{i}(r)=c_\mathrm{i}^2(r)/(\gamma-1)$ is the initial internal energy per unit mass. The case with $\beta=0$ represents locally isothermal disks, while the case with $\beta=\infty$ represents adiabatic disks. In this study, we take $\beta$ to be constant throughout the disk. 

The inner and outer boundaries of the computational domain are fixed to their initial values, while a periodic boundary condition is imposed on the azimuthal boundaries. To reduce wave reflection at the radial boundaries, we introduce wave-damping zones at $r_\mathrm{in} < r < 1.25r_\mathrm{in}$ and $0.8r_\mathrm{out} < r < r_\mathrm{out}$, following \citet{Val-Borro2006}. In these zones, the gas velocity is enforced to return to the initial velocity ${\bm v}_i$ on a timescale of $1/\eta$, according to 
\begin{equation}
  \frac{\partial {\bm v}}{\partial t} = -\eta ({\bm v} - {\bm v}_\mathrm{i}),
\end{equation}
where 
\begin{equation}
  \eta = \mathrm{min}(x,1)  \Omega_\mathrm{K} \\
\end{equation}
with
\begin{equation}
    x=
  \begin{cases}
    (1.25r_\mathrm{in}-r)/ r_\mathrm{p}, &  r_\mathrm{in}<r<1.25r_\mathrm{in},\\
    2.5(r-0.8r_\mathrm{out})/ r_\mathrm{p}, & 0.8r_\mathrm{out}<r<r_\mathrm{out}. 
  \end{cases} \\
\end{equation}

At the beginning of each simulation, 
the planet mass is gradually increased from zero to the final value $M_{\rm p}$ according to
\begin{equation}
  \tilde{M}_\mathrm{p}(t) =
  \begin{cases}
    M_\mathrm{p}\sin^2{\left[\frac{1}{2}\pi\left(\frac{t}{\tau_\mathrm{ins}}\right)\right]}, & t < \tau_\mathrm{ins},\\
    M_\mathrm{p},  &  t \geq \tau_\mathrm{ins}, 
  \end{cases} \\
\end{equation}
where $\tilde{M}_\mathrm{p}(t)$ is the planetary mass at time $t$. We set $\tau_\mathrm{ins} = 50 \times 2\pi/\Omega_\mathrm{K}(r_\mathrm{p})$.  

{Our simulations adopt code units such that $r_{\rm p} = \Omega_{\rm p} = \Sigma_{\rm 0} = 1$, where $\Omega_{\rm p} \equiv \Omega_\mathrm{K}(r_\mathrm{p})$ is the orbital frequency of the planet.
Since we take $Q_{\rm relax}$ to be proportional to $\Sigma$, equations (\ref{eq:EoC}), (\ref{eq:EoM}), and (\ref{eq:EoE}) are invariant under scaling $\Sigma \to C \Sigma$, where $C$ is an arbitrary constant. Therefore, the scaling factor $\Sigma_0$ in equation~\eqref{eq:3_s} can be chosen arbitrarily in our simulations. }

\subsection{Analysis of shock wave heating}
\label{subsec:analysis}

A planet embedded in a disk excites spiral density waves that travel through the disk away from the planet and eventually shock \citep{GoodmanRafikov2001, Rafikov2002}.
Because these waves rotate at the planet's orbital frequency, while the disk gas rotates differentially, they sweep through the disk gas at a given orbital radius $r$ with a frequency of $|\Omega_{\rm K}(r) - \Omega_{\rm p}|$. In this study, we quantify the shock wave heating rate by measuring the jump in the disk gas entropy across the planet-induced spiral waves.

The entropy change is a key physical quantity that characterizes the magnitude of gas heating at a shock. We denote the pre-shock gas density by $\Sigma_1$, pressure by $p_1$, and specific entropy by $s_1 = p_1/\Sigma_1^\gamma$, and the post-shock values by $\Sigma_2(>\Sigma_1)$, $p_2(>p_1)$, and $s_2= p_2/\Sigma_2^\gamma (>s_1)$, respectively. 
We define the dimensionless specific entropy jump $\delta$ as \begin{equation}
\delta \equiv \frac{s_2}{s_1} - 1 = \frac{p_2/\Sigma_2^\gamma}{p_1/\Sigma_1^\gamma}-1.
\label{eq:delta}
\end{equation}

\begin{figure*}
\begin{center}
\includegraphics[width=\linewidth, bb=0 0 782 504]{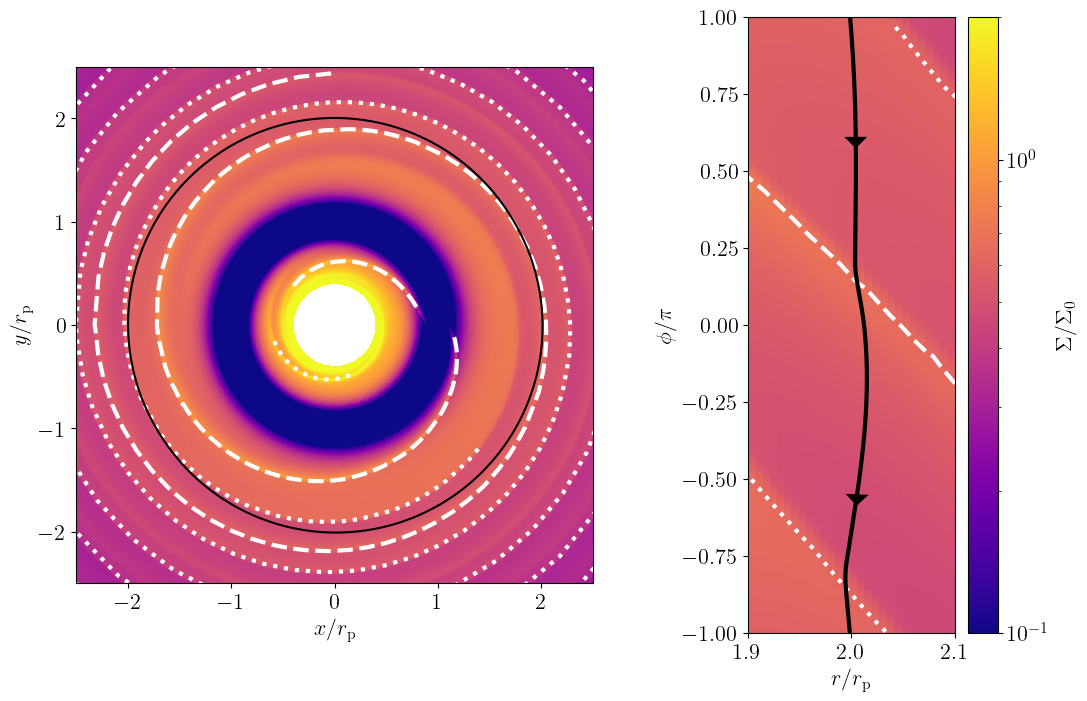}
\end{center}
\caption{Left panel: Snapshot of the 2D disk surface density distribution in the quasi-steady state from RUN 1. The white dashed and dotted lines mark the primary and secondary spiral arms, respectively. The black line indicates a streamline passing through the point $(r, \phi) = (2r_\mathrm{p}, 0)$. Right panel: Zoom-in of the vicinity of the streamline shown in the left panel. In the planet's rest frame, the gas flow on the streamline is in the direction of decreasing $\phi$.
} 
\label{f:gaikan}
\end{figure*}

The specific entropy jump is directly measurable from numerical simulations by tracing the motion of gas fluid elements. We take a snapshot of a simulation after the system reaches a quasi-steady state, which occurs after 1000 orbits of the planet for models with $\alpha = 10^{-3}$ and 3000 orbits for those with $\alpha = 10^{-3.5}$ or $10^{-4}$. On the snapshot, we draw gas streamlines originating from different orbital distances (see the right panel of figure~\ref{f:ryusen} in section~\ref{sec:kekka} for an example of such streamlines) and measure the entropy along each of them. {The entropy along the streamlines is calculated using bilinear interpolation of the 2D grid data}. Shocks manifest as jumps in the entropy profile. We calculate the $\delta$ value of each shock from the minimum and maximum entropy values immediately before and after the shock. By compiling the specific entropy jumps from different streamlines, we obtain $\delta$ as a function of $r$.
We discard small entropy jumps where $\delta$ is less than 20\% of the largest shock's value.

Our analysis excludes the region in the vicinity of the planet, specifically $0.8r_\mathrm{p} < r < 1.2r_\mathrm{p}$, where a deep planet-induced gap prevents reliable measurements of entropy changes associated with shocks. In practice, shock-induced heating is not significant in a deep planet-induced gap where the optical depth is small and radiative cooling is rapid \citep[see][]{Ziampras2020}.

The specific entropy jump can be translated into the heating rate $Q_{\mathrm{sh}}$ of the gas per unit area by \citep[see][]{Rafikov2016}
\begin{equation}
    Q_\mathrm{sh} = \frac{p_1}{\Delta t(\gamma-1)}\delta, \label{eq:Qsh}
\end{equation}
where $\Delta t = 2\pi/|\Omega_\mathrm{K}(r) - \Omega_\mathrm{p}|$ is the time interval between two consecutive shock crossings for a gas particle orbiting at radius $r$. As we show in the following section, the planet in our simulations induces multiple spirals both inside and outside of its orbit, with each spiral forming a shock. When multiple spiral arms exist along a single streamline, the net heating rate is proportional to the sum of the $\delta$ values at all these arms. {Using equation \eqref{eq:3_p}, the shock heating rate can also be written as}
\begin{equation}
Q_{\rm sh} = \frac{\Sigma c_{\rm s}^2}{\Delta t(\gamma-1)}\delta = \frac{\Sigma e}{2\pi}|\Omega_{\rm K}-\Omega_{\rm p}|\delta,
\label{eq:Qsh2}
\end{equation}
where we have omitted the subscript 1 for the preshock quantities. When there are multiple planet-induced shocks, $\delta$ in equation~\eqref{eq:Qsh2} should be regarded as the sum of the specific entropy jumps at the individual shocks.
{In section~\ref{subsec:overview}, we numerically confirm that equation~\eqref{eq:Qsh2} accurately describes the azimuthally averaged rate of heat generation by the spiral shocks.}

\subsection{Parameter sets}
\label{subsec:parameters}
We present {11} runs with different values of $(q, \xi, \alpha, \beta)$ as listed in table~\ref{table:param}, where $q={M}_{\rm p}/M_\ast$ is the planet-to-star mass ratio.  
We refer to RUN 1 as the fiducial model. 
{For all runs except RUN 6, the planet mass mass is larger than the thermal mass $q_{\rm th} = (c_0/v_{\rm K}(r_{\rm p}))^3 \approx 10^{-4}$, which is the minimum planet mass required for planet-induced waves to be strongly nonlinear and form a shock immediately after launching \citep{LinPapaloizou93,Rafikov2002}.}
Our simulations represent moderately adiabatic cases with $10 \leq \beta \leq 10^{2.5}$. For $\beta = 1$, we cannot accurately measure the $\delta$ values at the shocks because the strong cooling largely erases the entropy jumps. However, this is not a significant issue, as shock heating would have little effect on the disk temperature in such rapidly cooling disks \citep[see also][]{Ziampras2020}.
{In more adiabatic cases with $\beta > 10^{2}$, the disk's specific internal energy (or temperature) can significantly exceed the initial value, leading to numerical instability at the inner boundary, where the specific internal energy is fixed to the initial value. Our simulation is unstable with $\alpha = 10^{-3}$ and $\beta  \geq 10^{2.5}$, but is barely stable with $\alpha = 10^{-3.5}$ and $\beta = 10^{2.5}$ (RUN 11)}.

\begin{table}[t]
    \caption{List of parameter sets}
    \label{table:param}
    \centering
    \begin{tabular}{lcccc}
    \hline
    RUN  & $q$ & $\xi$ & $\alpha$ & $\beta$\\
    \hline 
    RUN 1 & $10^{-3}$ & $-1.0$ & $10^{-3}$ & $10^{2}$\\
    RUN 2 & $10^{-3}$ & $-0.5$ & $10^{-3}$ & $10^{2}$\\
    RUN 3 & $10^{-3}$ & $-1.5$ & $10^{-3}$ & $10^{2}$\\
    RUN 4 & $10^{-3.5}$ & $-1.0$ & $10^{-3}$ & $10^{2}$\\
    RUN 5 & $10^{-4}$ & $-1.0$ & $10^{-3}$ & $10^{2}$\\
    RUN 6 & $10^{-4.5}$ & $-1.0$ & $10^{-3}$ & $10^{2}$\\ 
    RUN 7 & $10^{-3}$ & $-1.0$ & $10^{-3.5}$ & $10^{2}$\\ 
    RUN 8 & $10^{-3}$ & $-1.0$ & $10^{-4}$ & $10^{2}$\\
    RUN 9 & $10^{-3}$ & $-1.0$ & $10^{-3}$ & $10$\\
    RUN 10 & $10^{-3}$ & $-1.0$ & $10^{-3}$ & $10^{1.5}$\\
{    RUN 11} & $10^{-3}$ & $-1.0$ & $10^{-3.5}$ & $10^{2.5}$\\    
    \hline
    \end{tabular}
\end{table}

\section{Results}\label{sec:kekka}

\subsection{The fiducial case}\label{subsec:overview}

\begin{figure}[t]
\includegraphics[width=\linewidth, bb=0 0 595 443]{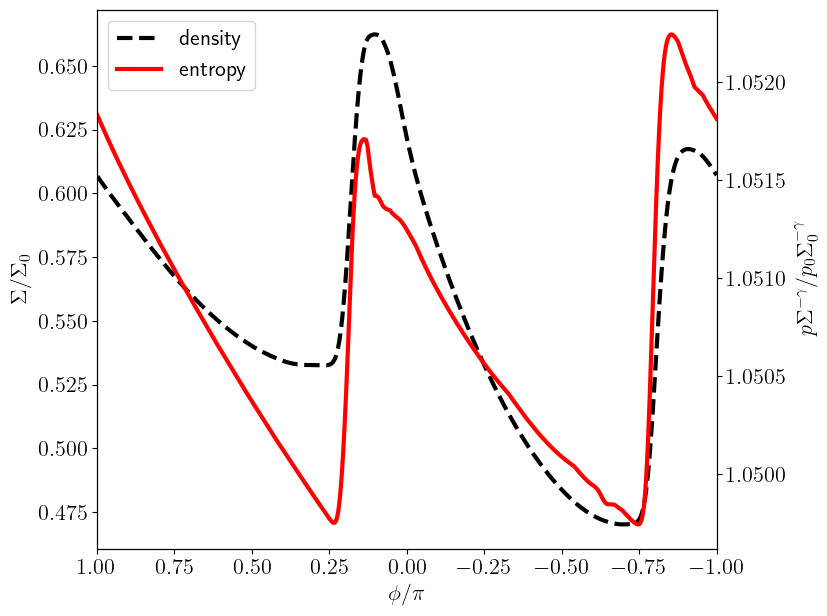}
    \caption{Distribution of the surface density (dashed line) and specific entropy $s = p\Sigma^{-\gamma}$ (red line) of the gas along the streamline shown in figure~\ref{f:gaikan}. The jumps in surface density and entropy observed at $\phi \sim 0.25\pi$ and $\sim -0.75\pi$ correspond to the shocks associated with the primary and secondary arms, respectively (see also the right panel of figure~\ref{f:gaikan}).
    }
    \label{f:ryusen}
\end{figure}

We begin by presenting the results from the fiducial run (RUN 1) to illustrate our shock wave analysis. The left panel of figure~\ref{f:gaikan} shows a snapshot of the 2D gas surface density profile in the quasi-steady state ($t=1000$ planetary orbits) from RUN 1. The planet creates a deep disk gap in its vicinity ($r \sim 0.8$--$1.2r_{\rm p}$). The planet also excites spiral waves both interior to and exterior to its orbit. As discussed in previous studies \citep{Zhu15,FungDong15,Bae17,Bae18a}, a planet excites more than one spiral arm in both the inner and outer disks.
Following the previous studies, we refer to the arms directly attached to the planet as the primary arms and to the strongest arms that do not connect to the planet as the secondary arms. In the left panel of figure~\ref{f:gaikan}, the primary and secondary arms are marked by the dashed and dotted lines, respectively.

As described in section~\ref{subsec:analysis}, we draw streamlines at different orbits and measure the entropy jumps across the arms. The black line in the left panel of figure \ref{f:gaikan} shows the streamline passing through the point $(r, \phi) = (2r_\mathrm{p}, 0)$ in this snapshot (see also the zoom-in of the vicinity of the streamline in the right panel of figure~\ref{f:gaikan}). Because the system has already reached a quasi-steady state, the streamlines on this snapshot approximately follow the trajectories of gas fluid elements. Figure~\ref{f:ryusen} shows how the gas surface density and entropy of a fluid element along this streamline vary in the direction of the flow in the planet's rest frame (negative $\phi$). The surface density of the element abruptly increases at the shock due to shock compression and then returns to the background value due to expansion. The entropy also increases at the shocks due to shock dissipation and then decreases due to cooling relaxation (the $Q_{\rm relax}$ term in equation~\eqref{eq:EoE}). After shock crossing, the change in entropy of the moving parcel decays on the timescale of $\sim t_{c}$, which is approximately 100 times the local Keplerian time. {A close inspection of figure~\ref{f:ryusen} shows that the measured entropy profile exhibits a small bump at $\phi \sim 0.1\pi$. This is a numerical artifact that arises when the streamline crosses a radial cell boundary and enters an adjacent outer cell with slightly higher entropy. However, this artificial bump does not affect our measurement of the entropy jump at the primary arm's shock, located at $\phi/\pi \sim 0.2$.}

\begin{figure}[t]
    \includegraphics[width=\linewidth, bb=0 0 525 451]{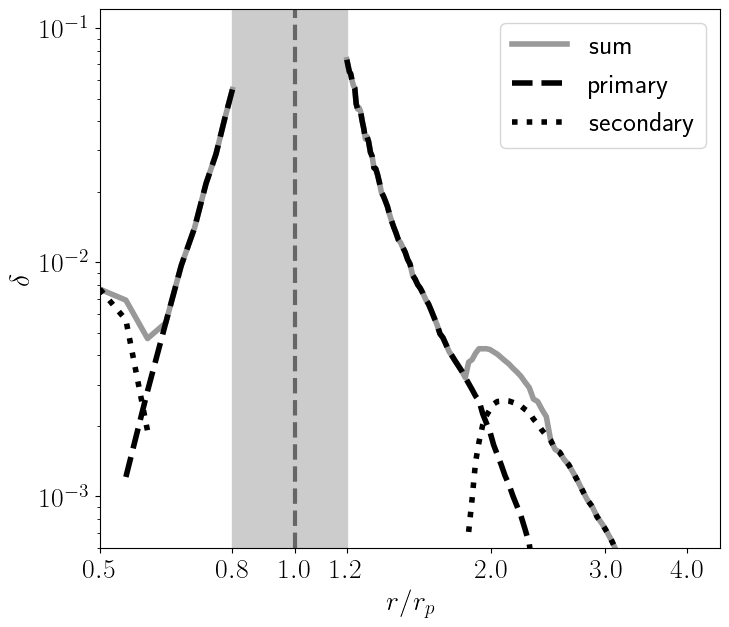}
    \caption{Radial profiles of the specific entropy jump $\delta$ (equation~\eqref{eq:delta}) for the primary and secondary arms (dashed and dotted lines, respectively) from RUN  1. The solid line shows the sum of the two jumps, representing the net shock heating rate. Our shock analysis excludes the shaded region of $0.8< r/r_\mathrm{p} < 1.2$, where the deep planet-carved gap prevents an accurate measurement of $\delta$ using streamlines.
    }
	\label{f:s_plot}
\end{figure}

The streamline analysis illustrated above provides the dimensionless specific entropy jumps $\delta$ (equation~\eqref{eq:delta}) at the primary and secondary shocks at different $r$, except in the gapped region $0.8 < r/r_{\rm p} < 1.2$, which is excluded from the analysis. These values are plotted in figure~\ref{f:s_plot} as a function of $r$. The primary inner and outer arms yield a non-zero $\delta$ at $r/r_{\rm p} = 0.8$ and 1.2, respectively, implying that both have already developed into shocks before propagating out of the gapped region. In contrast, the secondary inner and outer arms exhibit detectable entropy jumps at $r/r_{\rm p} \lesssim 0.7$ and $\gtrsim 2.0$, respectively, indicating that the secondary arms form shocks only after traveling considerable distances away from the planet. Interestingly, at large $|r - r_{\rm p}|$, the $\delta$ values of the secondary shocks surpass those of the primary shocks, meaning that disk heating by the secondary arms dominates over that by the primary ones. This is because the secondary arms have not experienced dissipation until they form shocks. The potential significance of the secondary shocks in disk heating is further discussed in section~\ref{subsec:heat}.

In the log--log plot of $\delta$ versus $r$ (figure~\ref{f:s_plot}), the radial profiles of $\delta$ for the inner and outer primary shocks appear as straight lines. This indicates that both approximately follow power laws. The same holds true for the fully developed secondary outer shock at $r/r_{\rm p} > 2$.
We cannot confirm such a trend for the secondary inner shock because the inner computational domain is too narrow for the secondary shock to fully develop.
We provide empirical formulas for the entropy jumps in section~\ref{subsec:empirical}.

\begin{figure}[t]
    \includegraphics[width=\linewidth, bb=0 0 472 413]{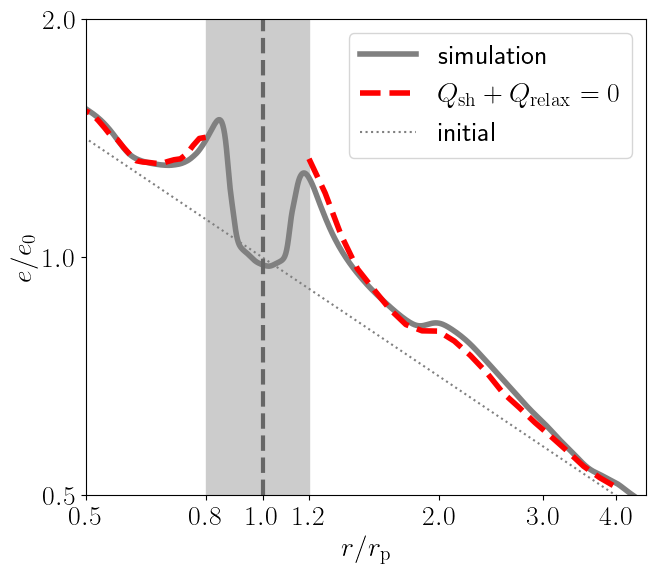}
    \caption{Radial profile of the azimuthally averaged specific internal energy $e$, normalized by $e_0 \equiv e_{\rm i}(r_{\rm p})$,  from RUN 1 in the quasi-steady state (solid line), compared with the prediction that assumes thermal equilibrium $Q_{\rm sh} + Q_{\rm relax} = 0$ (equation~\eqref{eq:e_equil}; dashed lines). The dotted line shows the initial profile $e = e_{\rm i}$. See the caption of figure~\ref{f:s_plot} for the shaded region. Note that viscous heating due to Keplerian shear is excluded from the simulation (see section~\ref{subsec:basiceqn}). 
    }
	\label{fig:e}
\end{figure}
{Before proceeding to a parameter study, we verify that the entropy jumps produced by the spiral shocks indeed cause disk heating, as predicted by equation~\eqref{eq:Qsh2}. Figure~\ref{fig:e} shows the radial profile of the azimuthally averaged specific internal energy from RUN 1 in the quasi-steady state. We expect that this internal energy profile is determined by the balance between shock heating and $\beta$ thermal relaxation (note that our simulations do not include viscous heating from Keplerian shear). Solving the  thermal equilibrium equation $Q_{\rm sh} + Q_{\rm relax} = 0$ for $e$, the internal energy profile in the steady state is predicted as
\begin{equation}
 e_{\rm equil} = e_{\rm i} \left( 1- \frac{\beta |1-\Omega_{\rm p}/\Omega(r)|}{2\pi}\delta\right)^{-1}.
 \label{eq:e_equil}
\end{equation}
We calculate $e_{\rm equil}$ using the radial $\delta$ profile obtained from the streamline analysis presented above. The result is overplotted in figure~\ref{fig:e}. We find that $e_{\rm equil}$ reproduces the azimuthally averaged $e$ in the steady state to within an accuracy of $\approx 5$\%.}

\begin{figure}[t]
    \includegraphics[width=\linewidth, bb=0 0 539 464]{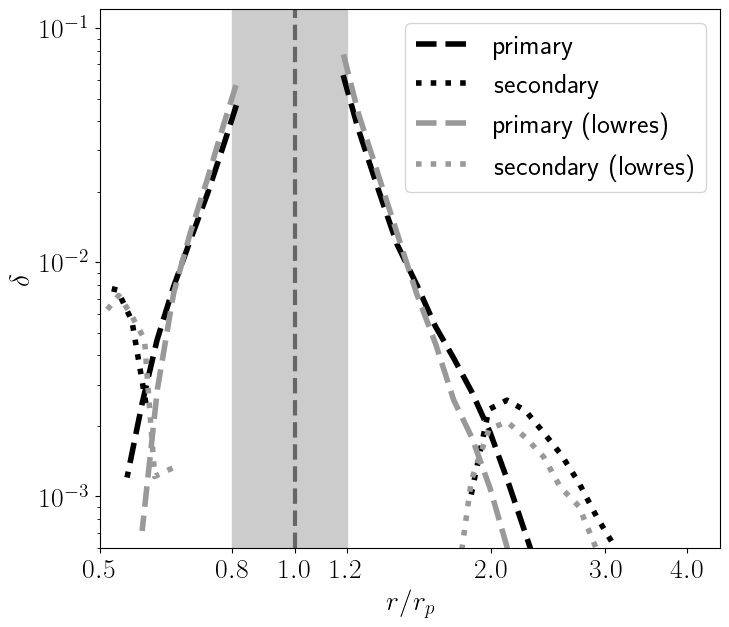}
    \caption{Radial profiles of the specific entropy jumps for the primary and secondary arms (dashed and dotted lines, respectively) from a low-resolution version of RUN 1 (gray lines), compared with those from RUN 1 at standard resolution shown in figure~\ref{f:s_plot} (black lines).    
    }
	\label{fig:resolution}
\end{figure}
{We also briefly examine the impact of numerical resolution on our measurements of entropy jumps. Here, we use a low-resolution version of RUN 1, with $448 \times 792$ cells, resulting in half the resolution in both the radial and azimuthal directions. Figure~\ref{fig:resolution} compares the radial profiles of $\delta$ for the primary and secondary arms from the low-resolution simulation with those from RUN 1 at standard resolution ($896\times1584$ cells). We find that the low-resolution run reproduces the $\delta$ values for relatively large jumps ($\delta \gtrsim 0.005$) within 20\% accuracy. The low-resolution run underestimates the smaller entropy jump at the secondary outer arm, but the error remains within $\approx$ 30\%. Based on these comparisons, we  expect that the simulations presented in this study produce $\delta$ values with a resolution-related error of less than 20--30\%.}

\subsection{Parameter survey}\label{subsec:PS}

\begin{figure*}
    \begin{center}
    \includegraphics[width=0.9\linewidth, bb=0 0 981 950]{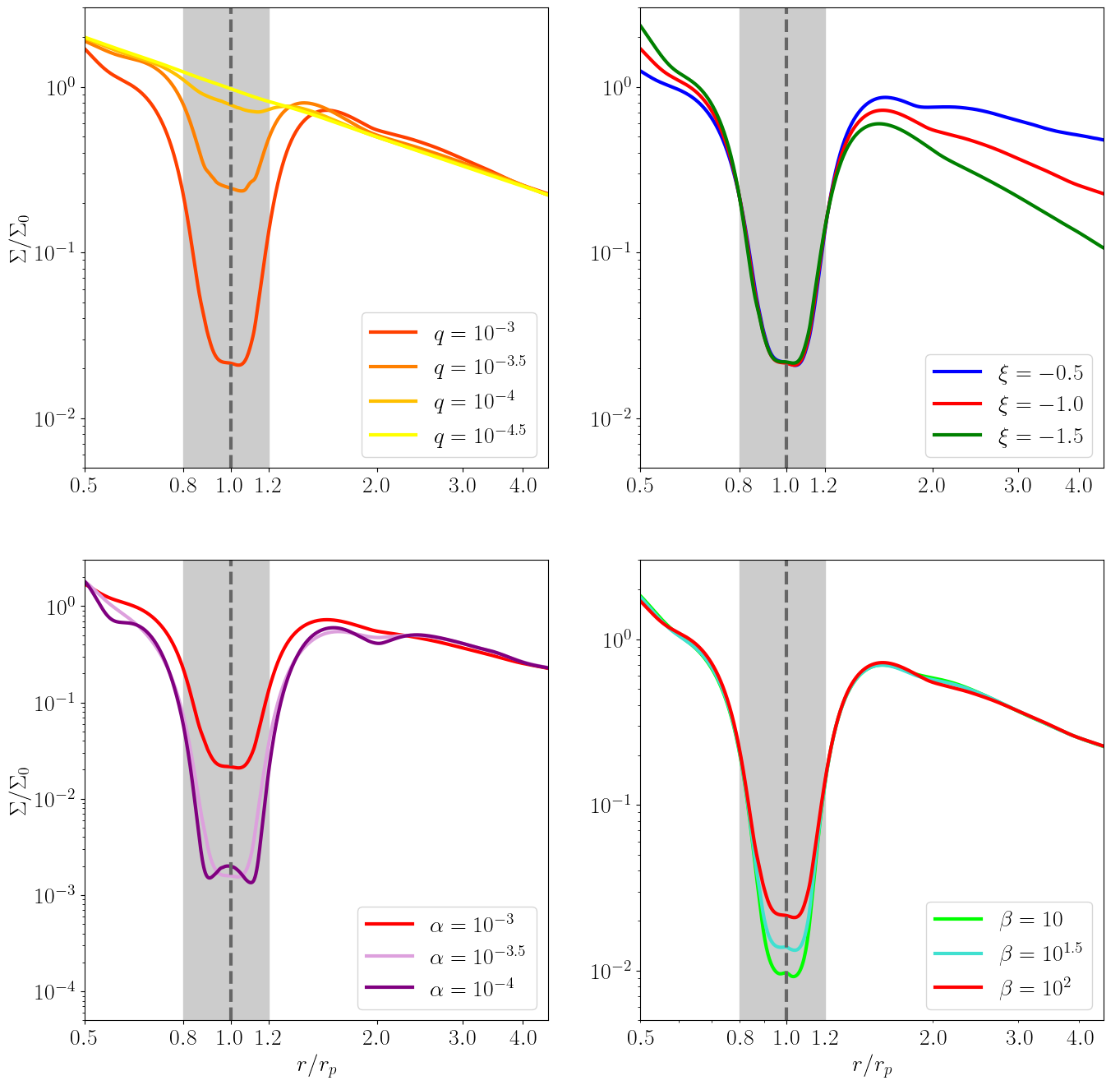}
    \end{center}
    \caption{Radial profiles of the azimuthally averaged surface density in the quasi-steady state from {RUNs 1--10}. The upper left, upper right, lower left, and lower right panels compare the results from runs with different planetary masses $q$ (RUNs 1 and 4--6), initial surface density slopes $\xi$ (RUNs 1--3), viscosity parameters $\alpha$ (RUNs 1, 7, and 8), and {thermal relaxation} timescales $\beta$ (RUNs 1,  9, 10), respectively.
       }
	\label{f:mitudo_plot}
\end{figure*}

We now explore how our simulation results depend on the model parameters. Figure~\ref{f:mitudo_plot} shows the radial profiles of the azimuthally averaged surface density in the quasi-steady state from {RUNs 1--10}. It can be seen that different values of the planet mass $q$, surface density slope $\xi$, viscosity parameter $\alpha$, and {thermal relaxation} rate $\beta$ result in varying surface density profiles. However, as we will see below, the radial profiles of the entropy jumps at the spiral shocks are surprisingly similar among the models when scaled properly.

\begin{figure*}
    \begin{center}
    \includegraphics[width=0.9\linewidth, bb=0 0 861 847]{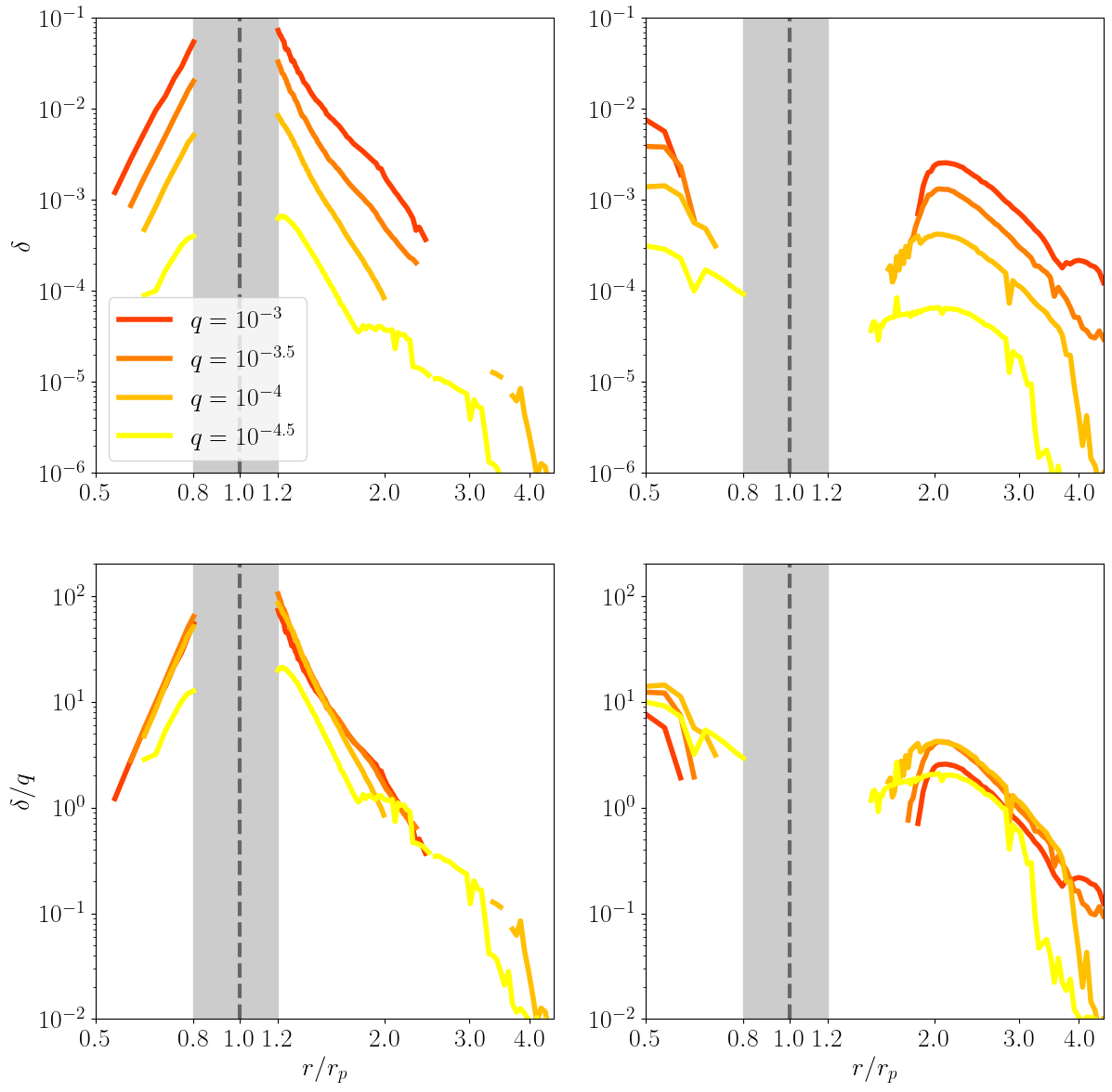}        
    \end{center}
    \caption{Upper panels: radial profiles of the specific entropy jumps $\delta$ at the primary and secondary arms (left and right panels, respectively) from runs with different planetary masses $q$ (RUNs 1 and 4--6). Lower panels: same as the upper panels, but show the entropy jumps normalized by the planetary mass, $\delta /q$.
    }
    \label{f:param_q}
\end{figure*}

The upper left and right panels of figure~\ref{f:param_q} show the radial profiles of $\delta$ for the primary and secondary arms, respectively, from runs with different planet masses (RUNs 1 and 4--6). Their behavior is qualitatively similar to that observed in RUN  1 (see the previous subsection). Specifically, regardless of the planet's mass, the $\delta$ values of the primary arms outside the gap region decreases monotonically with increasing $|r-r_{\rm p}|$, while the $\delta$ values of the secondary arms peak after traveling some distance from the planet. Once the secondary shock fully develops, its $\delta$ value exceeds that of the primary shock at the same location. The only significant difference is that the magnitude of $\delta$ scales approximately linearly with $q$. 
To see this, in the lower panels of figure~\ref{f:param_q}, we plot the entropy jumps normalized by the planetary mass, $\delta/q$, as a function of $r$ from the four runs. We observe that the radial profiles of the normalized entropy jump almost coincide, particularly in the cases of $q = 10^{-4}$--$10^{-3}$. The trend shifts slightly in the lowest-mass case of $q=10^{-4.5}$, potentially because the shocks for this case are too weak to resolve accurately.

\begin{figure*}[t]
    \begin{center}
    \includegraphics[width=0.9\linewidth, bb=0 0 861 430]{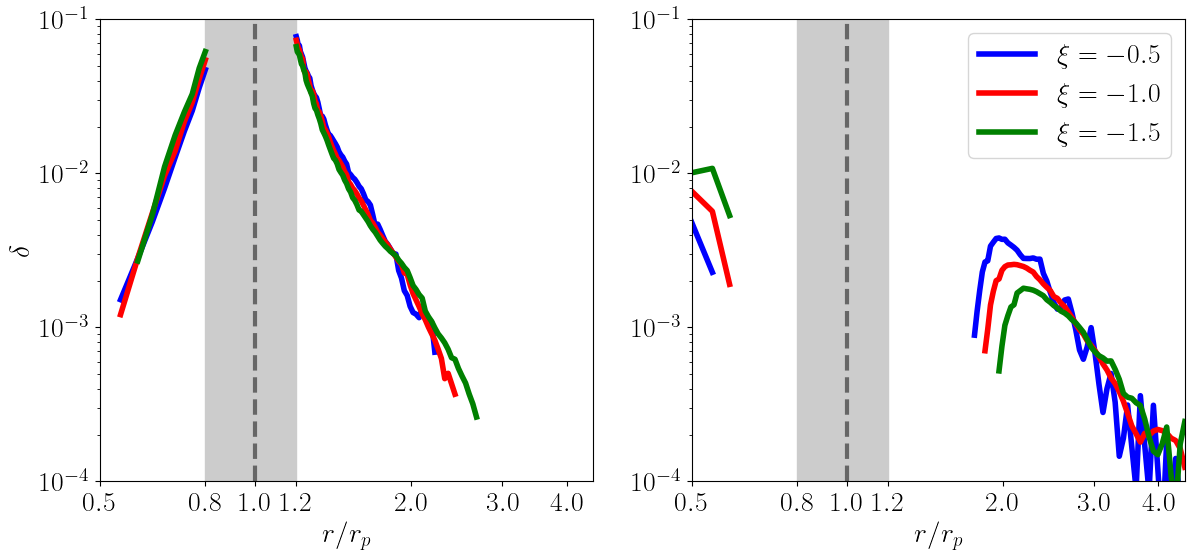}
    \end{center}
    \caption{Same as the upper panels of figure~\ref{f:param_q}, but from runs with different initial surface density slopes $\xi$ (RUNs 1--3).
    }
    \label{f:param_a}
\end{figure*}

Figure~\ref{f:param_a} shows the radial profiles of $\delta$ for the primary and secondary arms from runs with different values of $\xi$ (RUNs 1--3). The $\delta$ profile for the primary arms is nearly independent of $\xi$, despite the background surface density profile depending on $\xi$.
The same applies to the $\delta$ profile for the secondary arms. {The only notable trend is that the location where the secondary arms form a shock (i.e., where $\delta$ becomes detectable) depends slightly on $\xi$.  Specifically, as $\xi$ decreases, the shock formation position for the inner secondary arm shifts closer to the planet, while for the outer arm, it shifts farther. This dependence on $\xi$ likely reflects the initial radial sound velocity profile, $c_i \propto r^{-\xi-3/4}$ (see equation~\eqref{eq:3_c}). As $\xi$ decreases, the sound speed at $r < r_{\rm p}$ ($r > r_{\rm p}$) decreases (increases), causing the secondary arm to form a shock after traveling a shorter (longer) distance from the planet.}

\begin{figure*}
    \begin{center}
    \includegraphics[width=0.9\linewidth, bb=0 0 861 847]{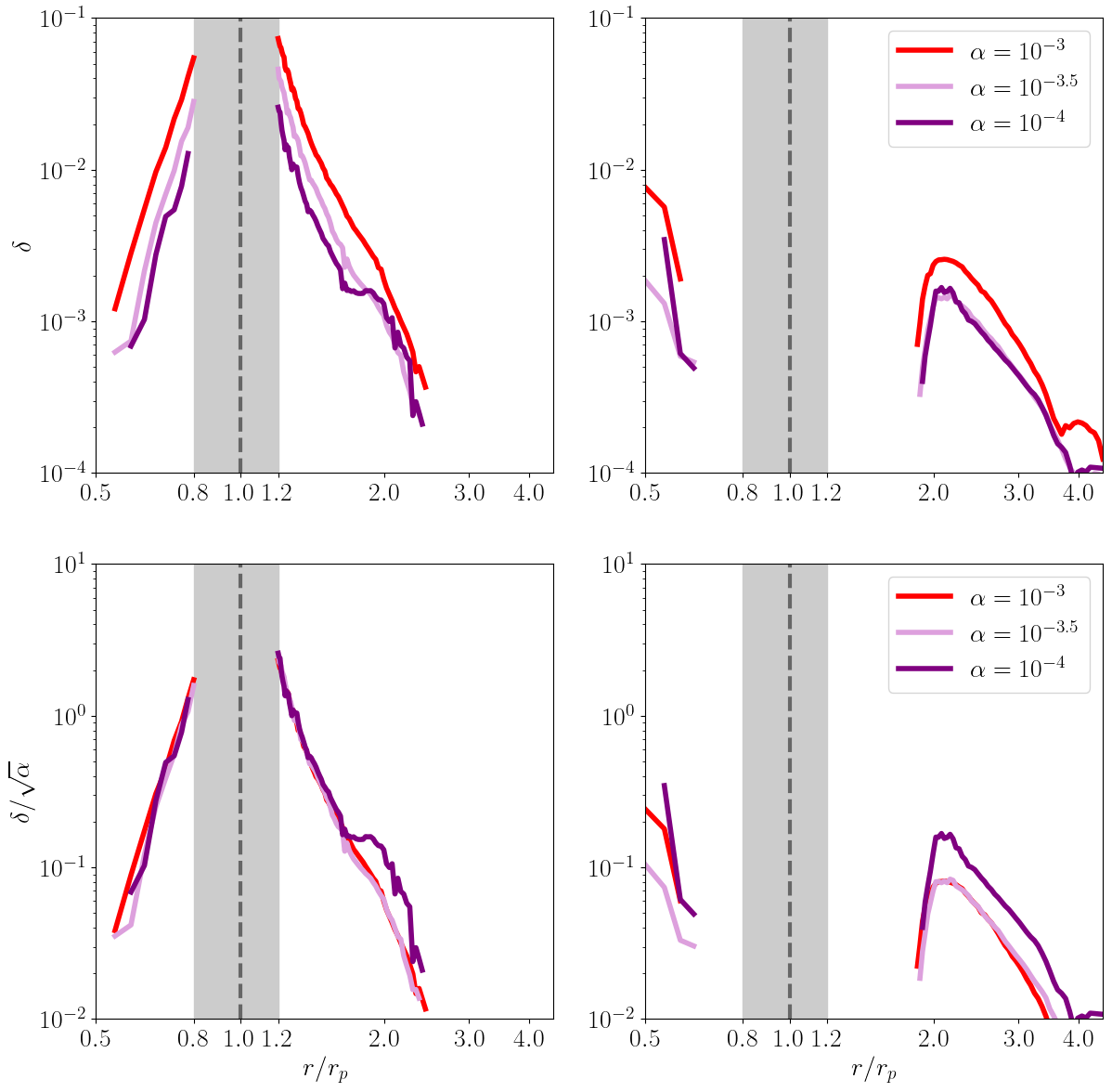}
    \end{center}
    \caption{Upper panels: Same as the upper panels of figure~\ref{f:param_q}, but from runs with different viscosity parameters $\alpha$ (RUNs 1, 7, and 8). Lower panels, same as the upper panels, but show the entropy jumps normalized by $\sqrt{\alpha}$.
     }
    \label{f:param_alp}
\end{figure*}

The upper panels of figure~\ref{f:param_alp} show the same as figure~\ref{f:param_a} but from runs with different values of $\alpha$ (RUNs 1, 7, and 8). In most cases, $\delta$ crudely scales as $\sqrt{\alpha}$, as demonstrated in the lower panels of figure~\ref{f:param_alp}. This indicates  that $\delta$ from our simulations depends weakly on the assumed viscosity. A notable exception is the secondary shock for  $\alpha=10^{-4}$, whose $\delta$ profile approximately follows that of the secondary shock for $\alpha = 10^{-3.5}$. 

\begin{figure*}[t]
    \begin{center}
    \includegraphics[width=0.9\linewidth, bb=0 0 861 430]{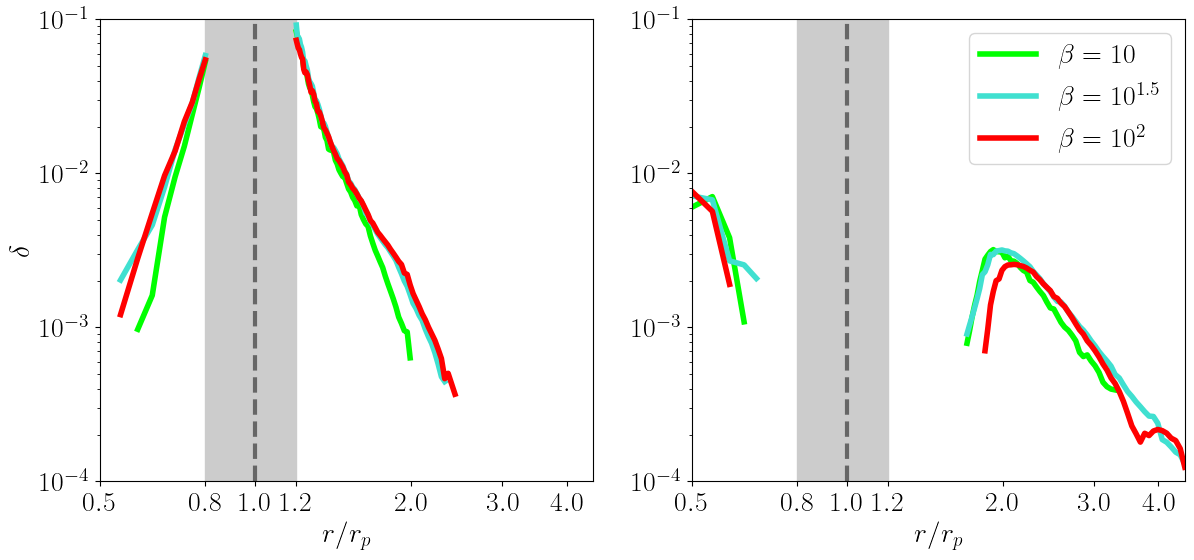}
    \end{center}
    \caption{Same as the upper panels of figure~\ref{f:param_q}, but from runs with different {thermal relaxation} timescales $\beta$ (RUNs 1, 9, and 10).
    }
    \label{f:param_beta}
\end{figure*}

Figure~\ref{f:param_beta} shows that varying $\beta$ over $10 \le \beta \le 10^2$ has little effect on $\delta$. {We note only that the peak of the $\delta$ distribution for the secondary outer arm shifts slightly outward as we vary $\beta$ from $10^{1.5}$ to $10^{2}$. This shift occurs because a higher $\beta$ results in an overall hotter disk with a higher sound speed, causing the secondary arm to form a shock farther from the planet.} 

\subsection{Empirical formulas for the entropy jumps}\label{subsec:empirical}

In the previous subsection, we have observed that $\delta$ for the primary and secondary arms follows universal relations that only depend on $q$, $\alpha$, and $r/r_{\rm p}$. Here, we provide empirical formulas for the universal relations.

\begin{figure*}[t]
  \begin{center}  
\includegraphics[width=0.7\linewidth, bb=0 0 598 611]{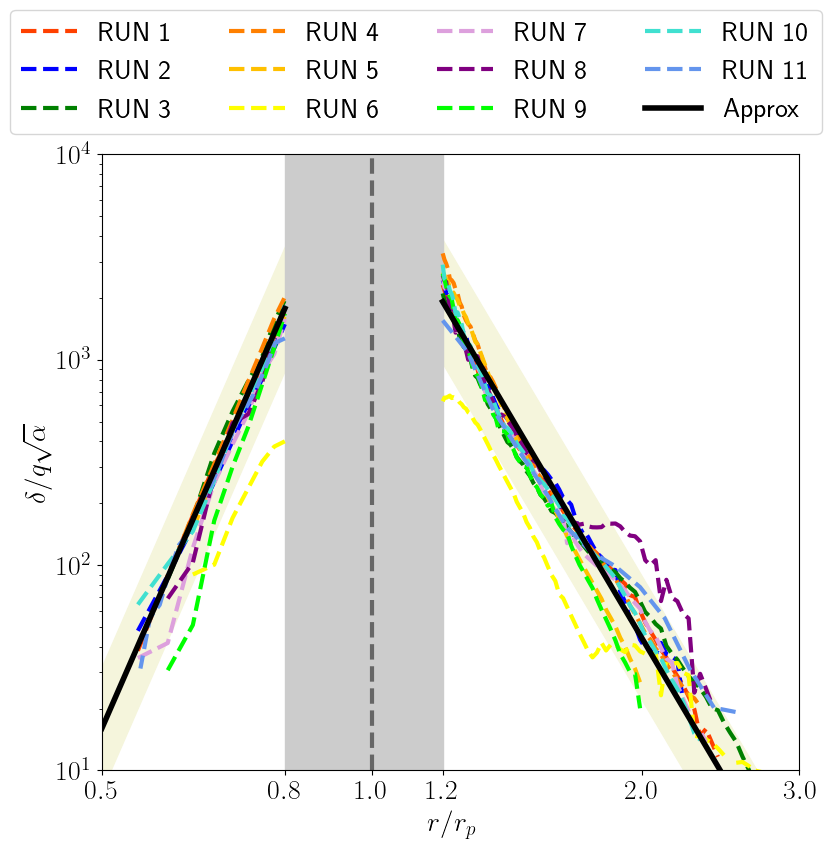}
  \end{center}  
    \caption{Universal relations for the specific entropy jumps for the primary arms. The dashed curves show the radial profiles of $\delta/q\sqrt{\alpha}$ for the primary inner and outer arms from all simulation runs. The black solid lines show the empirical formulas proposed in this study (equation~\eqref{eq:app_pri}). The yellow shading surrounding the solid lines indicates a factor of 2 deviation from the formulas. Our shock analysis excludes the gray-shaded region of $0.8< r/r_\mathrm{p} < 1.2$.
    }
    \label{f:app_pri}
\end{figure*}

For the primary arms, we assume that $\delta$ takes the power-law form
\begin{equation}
\delta=10^{f_0} q\sqrt{\alpha}\cdot|r/r_\mathrm{p}|^{f_1},
\label{eq:app_pri}
\end{equation}
where $f_0$ and $f_1$ are fitting parameters.
We searched for the set of $(f_0, f_1)$ that best reproduces the $\delta$ profiles for each of the primary inner and outer arms, using the data from all runs except for the lowest planet mass case (RUN 6; $q = 10^{-4.5}$) and the low viscosity case (RUN 8; $\alpha = 10^{-4}$), which appear to be slightly out of trend.
The obtained best-fit parameters for the primary inner and outer arms are $(f_0, f_1)=(4.22, 10.02)$ and $(f_0, f_1)=(3.87, -7.37)$,  respectively. 
Figure~\ref{f:app_pri} demonstrates that the best-fit formulas reproduce the results form all runs  except RUNs 6 and 8 to within a factor of 2.
Assuming radiative equilibrium where $Q_{\rm sh} \propto \sigma T^4$, with $\sigma$ being the Stefan--Boltzmann constant, a factor of 2 error in $\delta$ translates into an approximately 20\% error in the disk interior temperature $T$.
{For RUNs 6 and 8, our empirical formulas are less accurate, with a relative error of up to a factor of 3. The smaller-than-expected entropy jumps observed in RUN 6 are potentially due to the inefficient nonlinearity of the planet-induced waves, i.e., $q < q_{\rm th}$,  in this run (see section~\ref{subsec:parameters}). } 
{We note that the resolution-related errors in the measured $\delta$ values are less than 20--30\% (see section~\ref{subsec:overview}) and are therefore smaller than the errors arising from the fitting.}

\begin{figure}[t]
    \begin{center}
    \includegraphics[width=\linewidth, bb=0 0 520 540]{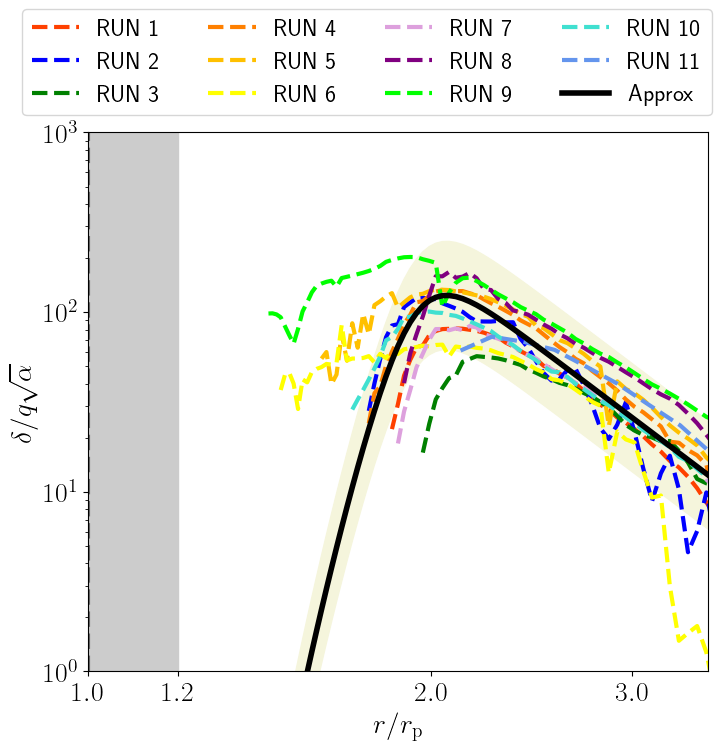}
    \end{center}
    \caption{Same as figure~\ref{f:app_pri}, but for the secondary outer arm.
    }
    \label{f:app_sec}
\end{figure}

We also provide an empirical formula for $\delta$ for the secondary outer arm, which has been found to fully develop into a shock within our computational domain. We assume that the secondary arm's $\delta$ also follows the form given by \eqref{eq:app_pri} but with a cutoff at small $|r - r_{\rm p}|$. By fitting our simulation results, we find that $\delta$ for the secondary outer arm scales as $r^{-4.79}$ at large radial distances. We propose the formula
\begin{equation}
\delta =10^{3.7}q\sqrt{\alpha}\pfrac{r}{r_\mathrm{p}}^{-4.79}\left[\pfrac{r/r_\mathrm{p}-1}{0.95}^{-12}+1\right]^{-1},
\label{eq:app_sec}
\end{equation}
where the last factor represents the cutoff.
This formula reproduces our simulation results to within a factor of a few (figure~\ref{f:app_sec}). 

\section{Discussion}\label{sec:giron}

\subsection{Comparison between the shock and viscous heating rates}\label{subsec:heat}

{As noted in section~\ref{subsec:basiceqn}, our simulations do not include viscous heating from the disk's Keplerian shear. In reality, however, both Keplerian shear and planet-induced shocks contribute to disk heating. Here, we use the analytic expressions for the specific entropy jump $\delta$ obtained in this study to predict when and where shock heating may dominate over viscous heating.}

 
The viscous heating rate is given by \citep{Lynden-BellPringle1974, Pringle1981}
\begin{equation}
Q_{\rm vis} = \frac{9}{4}\nu \Sigma \Omega_{\rm K}^2 
= \frac{9(\gamma-1)}{4}\alpha \Sigma e\Omega_{\rm K}.
\end{equation}
{The ratio between the shock heating rate $Q_{\rm sh}$ (equation~\eqref{eq:Qsh2}) and $Q_{\rm vis}$} can be written as  
\begin{eqnarray}
\frac{Q_{\rm sh}}{Q_{\rm vis}} &=& \frac{2}{9\pi (\gamma-1)}\left|1-\frac{\Omega_{\rm p}}{\Omega_{\rm K}(r)}\right| \frac{\delta}{\alpha}
\nonumber \\
&\approx& 0.18 \left|1-\left(\frac{r}{r_{\rm p}}\right)^{3/2}\right| \frac{\delta}{\alpha}.
\label{eq:Qratio}
\end{eqnarray}
Notably, the ratio depends only on $r/r_{\rm p}$ and $\delta/\alpha$. Furthermore, except at $r \gg r_{\rm p}$, the ratio is determined solely by $\delta/\alpha$. 

\begin{figure}[t]
    \includegraphics[width=\linewidth, bb=0 0 555 457]{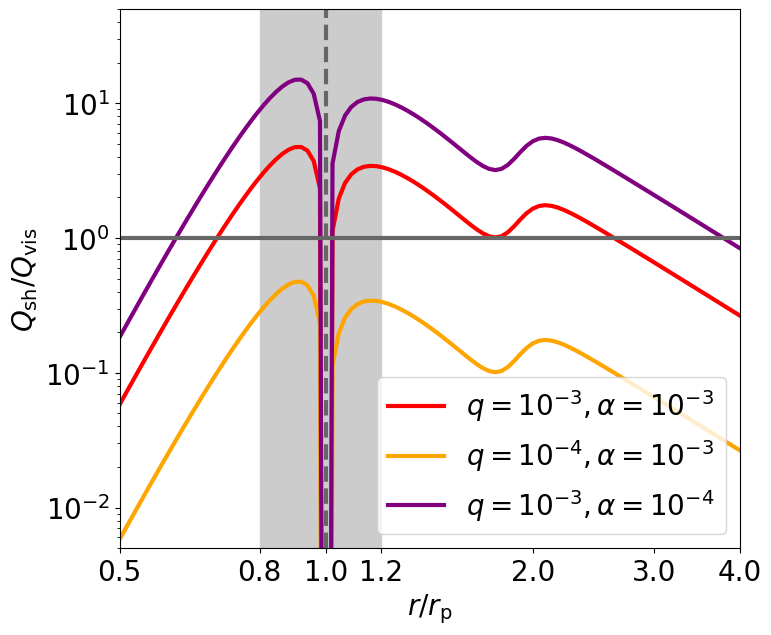}
    \caption{Ratio between the shock and viscous heating rates, $Q_{\rm sh}/Q_{\rm vis}$ (Equation~\eqref{eq:Qratio}), as a function of $r/r_{\rm p}$ for different values of the dimensionless planetary mass $q$ and viscosity $\alpha$. Here, the shock heating rate accounts for the primary and secondary shocks at $r > r_{\rm p}$ and only the primary shock at $r < r_{\rm p}$, using the empirical formulas for $\delta$ given by equations~\eqref{eq:app_pri} and \eqref{eq:app_sec}. Note that the formulas are validated only outside the gray-shaded region of $0.8< r/r_\mathrm{p} < 1.2$, which is excluded from our shock analysis.
    }
    \label{f:QvQ}
\end{figure}

Our empirically obtained relation $\delta \propto q\sqrt{\alpha}$  suggests that the heating rate ratio $Q_\mathrm{sh}/Q_\mathrm{vis}$ scales as $q/\sqrt{\alpha}$.  
Figure~\ref{f:QvQ} illustrates the radial distribution of
$Q_\mathrm{sh}/Q_\mathrm{vis}$ for $q = 10^{-4}$ and $10^{-3}$. Here, the shock heating rate accounts for  the contributions from the primary inner and outer arms, as well as the secondary outer arm (equations \eqref{f:app_pri} and \eqref{f:app_sec}, respectively).
We find that shock heating by a planet with mass $q = 10^{-3}$ can overwhelm viscous heating if $\alpha < 10^{-3}$. 
Our result aligns with the findings by \citet{Ziampras2020}, who showed that the shocks induced by a Jupiter-mass planet orbiting a solar-mass star ($q \approx 10^{-3}$) significantly heat a disk with $\alpha = 10^{-3}$ (see their figure 6). 

It is worth pointing out that the secondary outer shock produces a bump in $Q_{\rm sh}/Q_{\rm vis}$ at $r \approx 2 r_{\rm p}$. This bump is large enough for $Q_{\rm sh}$ to exceed $Q_{\rm vis}$ in the case of $q = 10^{-3}$ and $\alpha = 10^{-4}$. This suggests that the secondary arms may have a non-negligible impact on the disk's thermal structure away from the planetary orbit. This effect warrants further investigation in future hydrodynamical simulations with wider radial computational domains.

\subsection{Caveats and future work}\label{subsec:Scope}

We have demonstrated that shock heating by planet-induced spiral arms may serve as an important heating source for the disk harboring the planet. However, our simulations have two important limitations that should be addressed in future work.
Firstly, the computational domain employed in our simulations only covers $r > 0.5 r_{\rm p}$ due to computational cost. As a result, we could not fully observe how the secondary inner arm develops into a shock. Meanwhile, we have seen that the heating by the fully developed secondary outer shock can dominate over that by the primary outer shock, when compared at the same orbital radius. 
This is likely the case for the inner shocks as well, since we observe that the secondary shock's $\delta$ exceeds that of the primary shock already at the inner boundary of our computational domain (see, e.g., figure~\ref{f:s_plot}). 
Simulations with an extended inner computational domain, down to $r \sim 0.25 r_{\rm p}$, will allow us to confirm this expectation. Previous studies  \citep{Zhu15,Bae17} already showed that the tertiary inner arm can also emerge at $r \lesssim 0.5 r_{\rm p}$. Shock heating by this tertiary arm may potentially dominate over that by the secondary arm at large distances from the planet.

Secondly, our simulation models are 2D and therefore neglect any effects arising from the vertical structure of the disk and spiral arms.
The spiral shocks can be weaker in 3D disks, where the shocked gas is allowed to expand vertically, leading to adiabatic cooling \citep{Bate2003,BoleyDurisen2006,Lyra2016}.
Therefore, our 2D simulations could overestimate the magnitude of $\delta$. Nevertheless, it is possible that the scaling of $\delta$ with the model parameters and $r$ found in this study also holds in 3D disks. If this is the case, our 2D approximation would primarily affect the numerical prefactors in our empirical formulas for $\delta$, equations~\eqref{eq:app_pri} and \eqref{eq:app_sec}. 
In any case, 3D simulations will be necessary to validate our empirical formulas quantitatively. 

The simple $\beta$-{relaxation} prescription employed in this study may affect our disk's thermal structure. However, we expect that this treatment primarily influences disk cooling and does not significantly impact the  {\it shock heating} rate.
In fact, our simulations show that varying the dimensionless {thermal relaxation} timescale $\beta$ over the range $10 \leq \beta \leq 10^{2.5}$ has little effect on $\delta$ (see figure~\ref{f:param_beta}), indicating that the shock heating strength is insensitive to the {thermal relaxation} process as long as the {relaxation} timescale is much longer than the local orbital timescale. {For smaller $\beta$ ($\sim 0.01$--$1$), radiative damping of planet-induced spiral shocks \citep{Miranda2020} could affect $\delta$ profiles. However, we have not explored this parameter range as shock heating would have little impact on disk temperature in such rapidly cooling disks.}

\section{Conclusion}\label{sec:Sumally}

We performed 2D viscous hydrodynamical simulations of protoplanetary disks with an embedded planet to investigate disk heating caused by the shocks associated with planet-induced spiral density waves. We quantified the shock heating rate produced by the primary and secondary spiral arms by directly measuring the specific entropy jumps, $\delta$, across the shocks. We found that the radial profiles of $\delta$ approximately follow universal scaling relations that depend only on the planet-to-star mass ratio $q$, the dimensionless disk viscosity $\alpha$, and the normalized orbital radius (figures~\ref{f:app_pri} and \ref{f:app_sec}). Notably, $\delta$ is independent of the radial distributions of the disk's surface density, sound velocity, and {thermal relaxation} timescale, as long as the entropy jumps are measurable. We provided analytic expressions for $\delta$ for the primary inner and outer arms (equation~\eqref{eq:app_pri}), as well as for the secondary outer arm (equation~\eqref{eq:app_sec}). {These expressions are accurate within a factor of two for a moderately viscous ($10^{-3.5} \lesssim \alpha \lesssim  10^{-3}$) and moderately adiabatic ($10 \lesssim  \beta \lesssim  10^{2.5}$) disk with a planet massive enough that its spiral arms are strongly nonlinear ($q \gtrsim q_{\rm th}$). For cases with lower viscosity or a smaller planet, the expressions are less accurate, with deviations up to a factor of three for $q = 10^{-4.5}$ and $\alpha = 10^{-4}$ (RUNs 6 and 8, respectively).}

The empirically obtained expressions for the shock heating rate enables one to predict the shock and viscous heating rates for various values of $q$ and $\alpha$. We found that shock heating can dominate over viscous heating when $\alpha \lesssim 10^{-3}$ and $q \gtrsim 10^{-3}$ (figure \ref{f:QvQ}). Our formulas can also be easily implemented into radially one-dimensional models of gas and dust evolution \citep[e.g.,][]{Oka_2011, Bitsch2015, Delage23} and will thus enable the study of how planet-induced shock heating affects gas and dust evolution, as well as planet formation, after a massive planet has formed.
 
Our results are based on 2D simulations with a relatively narrow inner computational domain. Simulations with an extended radial domain are necessary to model shock heating by the secondary and/or tertiary inner arms. Additionally, 3D simulations are needed to quantify the effects of vertical dynamics on the shock heating rate.


\begin{ack}
We thank Hidekazu Tanaka for useful discussions, Ryosuke Tominaga for useful comments on the treatment of viscous heating in the {\tt Athena++} code, and the anonymous referee for insightful and constructive comments that helped to improve the quality of this work.
\end{ack}

\section*{Funding}
This work was supported by JSPS KAKENHI Grant Numbers JP20H00182, JP23H01227, JP23K25923, and JP23K03463.

\section*{Data availability} 
The data underlying this article will be shared on reasonable request
to the corresponding author.






\bibliographystyle{style/apj}
\bibliography{reference}

\end{document}